\documentclass[twocolumn,letterpaper,aps,prl,longbibliography,superscriptaddress,nofootinbib,showpacs,floatfix]{revtex4-1}

\usepackage{graphicx}   

\usepackage{xspace} 
\usepackage{amsmath}    

\newcommand{\sqsn}{\mbox{$\sqrt{s_{_{NN}}}$}\xspace}

\begin{document}

\title{Measurement of long-range angular correlation 
and quadrupole anisotropy of pions and (anti)protons 
in central $d$$+$Au collisions at $\sqrt{s_{_{NN}}}$~=~200~GeV}

\newcommand{\abilene}{Abilene Christian University, Abilene, Texas 79699, USA}
\newcommand{\augie}{Department of Physics, Augustana College, Sioux Falls, South Dakota 57197, USA}
\newcommand{\banaras}{Department of Physics, Banaras Hindu University, Varanasi 221005, India}
\newcommand{\barc}{Bhabha Atomic Research Centre, Bombay 400 085, India}
\newcommand{\baruch}{Baruch College, City University of New York, New York, New York, 10010 USA}
\newcommand{\bnlcoll}{Collider-Accelerator Department, Brookhaven National Laboratory, Upton, New York 11973-5000, USA}
\newcommand{\bnlphys}{Physics Department, Brookhaven National Laboratory, Upton, New York 11973-5000, USA}
\newcommand{\caucr}{University of California - Riverside, Riverside, California 92521, USA}
\newcommand{\charlesczech}{Charles University, Ovocn\'{y} trh 5, Praha 1, 116 36, Prague, Czech Republic}
\newcommand{\chonbuk}{Chonbuk National University, Jeonju, 561-756, Korea}
\newcommand{\ciae}{Science and Technology on Nuclear Data Laboratory, China Institute of Atomic Energy, Beijing 102413, P.~R.~China}
\newcommand{\cns}{Center for Nuclear Study, Graduate School of Science, University of Tokyo, 7-3-1 Hongo, Bunkyo, Tokyo 113-0033, Japan}
\newcommand{\colorado}{University of Colorado, Boulder, Colorado 80309, USA}
\newcommand{\columbia}{Columbia University, New York, New York 10027 and Nevis Laboratories, Irvington, New York 10533, USA}
\newcommand{\czechtech}{Czech Technical University, Zikova 4, 166 36 Prague 6, Czech Republic}
\newcommand{\dapnia}{Dapnia, CEA Saclay, F-91191, Gif-sur-Yvette, France}
\newcommand{\elte}{ELTE, E{\"o}tv{\"o}s Lor{\'a}nd University, H - 1117 Budapest, P{\'a}zm{\'a}ny P. s. 1/A, Hungary}
\newcommand{\ewha}{Ewha Womans University, Seoul 120-750, Korea}
\newcommand{\fit}{Florida Institute of Technology, Melbourne, Florida 32901, USA}
\newcommand{\fsu}{Florida State University, Tallahassee, Florida 32306, USA}
\newcommand{\gsu}{Georgia State University, Atlanta, Georgia 30303, USA}
\newcommand{\hanyang}{Hanyang University, Seoul 133-792, Korea}
\newcommand{\hiroshima}{Hiroshima University, Kagamiyama, Higashi-Hiroshima 739-8526, Japan}
\newcommand{\ihepprot}{IHEP Protvino, State Research Center of Russian Federation, Institute for High Energy Physics, Protvino, 142281, Russia}
\newcommand{\illuiuc}{University of Illinois at Urbana-Champaign, Urbana, Illinois 61801, USA}
\newcommand{\inrras}{Institute for Nuclear Research of the Russian Academy of Sciences, prospekt 60-letiya Oktyabrya 7a, Moscow 117312, Russia}
\newcommand{\instpasczech}{Institute of Physics, Academy of Sciences of the Czech Republic, Na Slovance 2, 182 21 Prague 8, Czech Republic}
\newcommand{\isu}{Iowa State University, Ames, Iowa 50011, USA}
\newcommand{\jaea}{Advanced Science Research Center, Japan Atomic Energy Agency, 2-4 Shirakata Shirane, Tokai-mura, Naka-gun, Ibaraki-ken 319-1195, Japan}
\newcommand{\jyvaskyla}{Helsinki Institute of Physics and University of Jyv{\"a}skyl{\"a}, P.O.Box 35, FI-40014 Jyv{\"a}skyl{\"a}, Finland}
\newcommand{\kek}{KEK, High Energy Accelerator Research Organization, Tsukuba, Ibaraki 305-0801, Japan}
\newcommand{\korea}{Korea University, Seoul, 136-701, Korea}
\newcommand{\kurchatov}{Russian Research Center ``Kurchatov Institute", Moscow, 123098 Russia}
\newcommand{\kyoto}{Kyoto University, Kyoto 606-8502, Japan}
\newcommand{\labllr}{Laboratoire Leprince-Ringuet, Ecole Polytechnique, CNRS-IN2P3, Route de Saclay, F-91128, Palaiseau, France}
\newcommand{\lahorelums}{Physics Department, Lahore University of Management Sciences, Lahore, Pakistan}
\newcommand{\lawllnl}{Lawrence Livermore National Laboratory, Livermore, California 94550, USA}
\newcommand{\losalamos}{Los Alamos National Laboratory, Los Alamos, New Mexico 87545, USA}
\newcommand{\lpc}{LPC, Universit{\'e} Blaise Pascal, CNRS-IN2P3, Clermont-Fd, 63177 Aubiere Cedex, France}
\newcommand{\lund}{Department of Physics, Lund University, Box 118, SE-221 00 Lund, Sweden}
\newcommand{\maryland}{University of Maryland, College Park, Maryland 20742, USA}
\newcommand{\mass}{Department of Physics, University of Massachusetts, Amherst, Massachusetts 01003-9337, USA }
\newcommand{\michigan}{Department of Physics, University of Michigan, Ann Arbor, Michigan 48109-1040, USA}
\newcommand{\muenster}{Institut fur Kernphysik, University of Muenster, D-48149 Muenster, Germany}
\newcommand{\muhlenberg}{Muhlenberg College, Allentown, Pennsylvania 18104-5586, USA}
\newcommand{\myongji}{Myongji University, Yongin, Kyonggido 449-728, Korea}
\newcommand{\nagasaki}{Nagasaki Institute of Applied Science, Nagasaki-shi, Nagasaki 851-0193, Japan}
\newcommand{\natmephi}{National Research Nuclear University, MEPhI, Moscow Engineering Physics Institute, Moscow, 115409, Russia}
\newcommand{\newmex}{University of New Mexico, Albuquerque, New Mexico 87131, USA }
\newcommand{\nmsu}{New Mexico State University, Las Cruces, New Mexico 88003, USA}
\newcommand{\ohio}{Department of Physics and Astronomy, Ohio University, Athens, Ohio 45701, USA}
\newcommand{\ornl}{Oak Ridge National Laboratory, Oak Ridge, Tennessee 37831, USA}
\newcommand{\orsay}{IPN-Orsay, Universite Paris Sud, CNRS-IN2P3, BP1, F-91406, Orsay, France}
\newcommand{\peking}{Peking University, Beijing 100871, P.~R.~China}
\newcommand{\pnpi}{PNPI, Petersburg Nuclear Physics Institute, Gatchina, Leningrad region, 188300, Russia}
\newcommand{\riken}{RIKEN Nishina Center for Accelerator-Based Science, Wako, Saitama 351-0198, Japan}
\newcommand{\rikjrbrc}{RIKEN BNL Research Center, Brookhaven National Laboratory, Upton, New York 11973-5000, USA}
\newcommand{\rikkyo}{Physics Department, Rikkyo University, 3-34-1 Nishi-Ikebukuro, Toshima, Tokyo 171-8501, Japan}
\newcommand{\saopaulo}{Universidade de S{\~a}o Paulo, Instituto de F\'{\i}sica, Caixa Postal 66318, S{\~a}o Paulo CEP05315-970, Brazil}
\newcommand{\seoulnat}{Seoul National University, Seoul, Korea}
\newcommand{\stonybrkc}{Chemistry Department, Stony Brook University, SUNY, Stony Brook, New York 11794-3400, USA}
\newcommand{\stonycrkp}{Department of Physics and Astronomy, Stony Brook University, SUNY, Stony Brook, New York 11794-3800,, USA}
\newcommand{\tenn}{University of Tennessee, Knoxville, Tennessee 37996, USA}
\newcommand{\titech}{Department of Physics, Tokyo Institute of Technology, Oh-okayama, Meguro, Tokyo 152-8551, Japan}
\newcommand{\tsukuba}{Institute of Physics, University of Tsukuba, Tsukuba, Ibaraki 305, Japan}
\newcommand{\vandy}{Vanderbilt University, Nashville, Tennessee 37235, USA}
\newcommand{\waseda}{Waseda University, Advanced Research Institute for Science and Engineering, 17 Kikui-cho, Shinjuku-ku, Tokyo 162-0044, Japan}
\newcommand{\weizmann}{Weizmann Institute, Rehovot 76100, Israel}
\newcommand{\wigner}{Institute for Particle and Nuclear Physics, Wigner Research Centre for Physics, Hungarian Academy of Sciences (Wigner RCP, RMKI) H-1525 Budapest 114, POBox 49, Budapest, Hungary}
\newcommand{\yonsei}{Yonsei University, IPAP, Seoul 120-749, Korea}
\affiliation{\abilene}
\affiliation{\augie}
\affiliation{\banaras}
\affiliation{\barc}
\affiliation{\baruch}
\affiliation{\bnlcoll}
\affiliation{\bnlphys}
\affiliation{\caucr}
\affiliation{\charlesczech}
\affiliation{\chonbuk}
\affiliation{\ciae}
\affiliation{\cns}
\affiliation{\colorado}
\affiliation{\columbia}
\affiliation{\czechtech}
\affiliation{\dapnia}
\affiliation{\elte}
\affiliation{\ewha}
\affiliation{\fit}
\affiliation{\fsu}
\affiliation{\gsu}
\affiliation{\hanyang}
\affiliation{\hiroshima}
\affiliation{\ihepprot}
\affiliation{\illuiuc}
\affiliation{\inrras}
\affiliation{\instpasczech}
\affiliation{\isu}
\affiliation{\jaea}
\affiliation{\jyvaskyla}
\affiliation{\kek}
\affiliation{\korea}
\affiliation{\kurchatov}
\affiliation{\kyoto}
\affiliation{\labllr}
\affiliation{\lahorelums}
\affiliation{\lawllnl}
\affiliation{\losalamos}
\affiliation{\lpc}
\affiliation{\lund}
\affiliation{\maryland}
\affiliation{\mass}
\affiliation{\michigan}
\affiliation{\muenster}
\affiliation{\muhlenberg}
\affiliation{\myongji}
\affiliation{\nagasaki}
\affiliation{\natmephi}
\affiliation{\newmex}
\affiliation{\nmsu}
\affiliation{\ohio}
\affiliation{\ornl}
\affiliation{\orsay}
\affiliation{\peking}
\affiliation{\pnpi}
\affiliation{\riken}
\affiliation{\rikjrbrc}
\affiliation{\rikkyo}
\affiliation{\saopaulo}
\affiliation{\seoulnat}
\affiliation{\stonybrkc}
\affiliation{\stonycrkp}
\affiliation{\tenn}
\affiliation{\titech}
\affiliation{\tsukuba}
\affiliation{\vandy}
\affiliation{\waseda}
\affiliation{\weizmann}
\affiliation{\wigner}
\affiliation{\yonsei}
\author{A.~Adare} \affiliation{\colorado}
\author{C.~Aidala} \affiliation{\losalamos} \affiliation{\mass} \affiliation{\michigan}
\author{N.N.~Ajitanand} \affiliation{\stonybrkc}
\author{Y.~Akiba} \affiliation{\riken} \affiliation{\rikjrbrc}
\author{R.~Akimoto} \affiliation{\cns}
\author{H.~Al-Bataineh} \affiliation{\nmsu}
\author{H.~Al-Ta'ani} \affiliation{\nmsu}
\author{J.~Alexander} \affiliation{\stonybrkc}
\author{K.R.~Andrews} \affiliation{\abilene}
\author{A.~Angerami} \affiliation{\columbia}
% ...
\collaboration{PHENIX Collaboration} \noaffiliation

\date{\today}

%
%

\begin{abstract}


We present azimuthal angular correlations between charged hadrons
and energy deposited in calorimeter towers in central $d$$+$Au and
minimum bias $p$$+$$p$ collisions at $\sqrt{s_{_{NN}}}=200$~GeV. The 
charged hadron is measured at midrapidity $|\eta|<0.35$, and the energy
is measured at large rapidity ($-3.7<\eta<-3.1$, Au-going
direction). An enhanced near-side angular correlation across
$|\Delta\eta| >$ 2.75 is observed in $d$$+$Au collisions. Using the
event plane method applied to the Au-going energy distribution, we
extract the anisotropy strength $v_2$ for inclusive charged
hadrons at midrapidity up to $p_T=4.5$~GeV/$c$. We also present
the measurement of $v_2$ for identified $\pi^{\pm}$ and
(anti)protons in central $d$$+$Au collisions, and observe a
mass-ordering pattern similar to that seen in heavy ion
collisions. These results are compared with viscous hydrodynamic
calculations and measurements from $p$$+$Pb at 
$\sqrt{s_{_{NN}}}=5.02$~TeV. The
magnitude of the mass-ordering in $d$$+$Au is found to be smaller
than that in $p$$+$Pb collisions, which may indicate smaller radial
flow in lower energy $d$$+$Au collisions.
\end{abstract}

\pacs{25.75.Dw}

\maketitle


%
%

Small collision systems, $d$$+$Au and $p$$+$Pb, have been studied at 
the Relativistic Heavy Ion Collider (RHIC) and the Large Hadron Collider 
(LHC) to understand baseline nuclear effects for heavy-ion collisions in 
which hot nuclear matter is made. The $d$$+$Au and $p$$+$Pb systems have 
generally been considered too small to create significant quantities of 
hot nuclear matter. This assumption has been challenged in $p$$+$Pb at 
\sqsn~=~5.02~TeV with the measurements of (i) near-side azimuthal 
correlations across a large pseudorapidity gap~\cite{CMS:2012qk,Aad:2012,Abelev:2012cya}, also observed in high multiplicity $p$$+$$p$ collisions
at 7~TeV~\cite{Khachatryan:2010gv}, and (ii)the elliptic anisotropy 
parameter $v_2$ measured by multiple particle 
correlations~\cite{Atlas:2013cm,CMS:2013cm}. 

Hydrodynamic models, successfully applied to heavy ion data at RHIC
and the LHC, can qualitatively reproduce the $v_2$ results from 
$p$/$d$+nucleus~\cite{Bozek:2013uha,Bzdak:2013zma,Qin:2013bha}.
If hydrodynamics is the primary cause of the observed effects then there
should be a mass-ordering of the magnitudes of $v_2$ for identified
particles, in which heavier particles have smaller $v_2$ values at low
$p_T<1.5$~GeV/$c$~\cite{Huovinen:2001cy, Bozek:2013ska}. Recently, such
mass-ordering has been observed in $p$$+$Pb collisions at LHC for $v_2$ of
$\pi^{\pm}$ and $p,\bar{p}$~\cite{ABELEV:2013wsa}. Finite near-side 
correlations can also arise from enhanced two-gluon 
emission at high parton densities as in the Color Glass Condensate (CGC) 
model~\cite{Dusling:2012iga,Dusling:2012wy,Dusling:2013oia}. 

Long-range angular correlations and elliptic anisotropy of
inclusive and identified hadrons in $p$$+$$p$ and $d$$+$Au collisions
at RHIC can provide crucial tests as to whether a hydrodynamically
expanding medium is created in these small systems. The $v_2$ in $d$$+$Au has 
been measured from hadron pair correlations, within a limited
rapidity range ($0.7>|\Delta\eta|>$ 0.48) and under the assumption that 
jet-like correlations are the same in various multiplicity-selected events~\cite{Adare:2013piz}.
In this Letter, we report measurements of azimuthal correlations in top
5\% central $d$$+$Au and minimum bias $p$$+$$p$ collisions between charged
hadrons at midrapidity ($|\eta|<0.35$) and energy deposited at
large rapidity $-3.7<\eta<-3.1$ (Au-going direction). We also
report $v_2$ for inclusive hadron and identified pions and
(anti)protons in $d$$+$Au at midrapidity using an event plane
across $|\Delta\eta| >$ 2.75.

The data were obtained from $p$$+$$p$ in the 2008 and 2009 experimental 
runs and $d$$+$Au in the 2008 run with the PHENIX detector. The event 
centrality class in $d$$+$Au collisions is determined as a percentile of 
the total charge measured in the PHENIX beam-beam counter covering $-3.9 < 
\eta < -3.0$ on the Au-going side~\cite{Adare:2013pc,ppg146,Adare:2011sc,PhysRevLett.111.202301}. For the 5\% most 
central $d$$+$Au collisions, the corresponding number of binary collisions 
and number of participants are estimated by a Glauber model to be $18.1\pm1.2$ and $17.8\pm1.2$ respectively~\cite{Adare:2013pc}.

%
%


Charged particles used in this analysis are reconstructed in the two 
PHENIX central-arm tracking systems, consisting of drift chambers and 
multi-wire proportional pad chambers (PC)~\cite{Adcox:2003zm}. 
Each arm covers ${\pi}/2$ in azimuth and $|\eta|<0.35$, and the tracking
system achieves a momentum resolution of $\delta$$p/p$$\approx$0.7\%$\oplus$1.1\%$\times$$p$ (GeV/$c$).

The drift-chamber tracks are matched to hits in the third layer 
of the PC, reducing the contribution of tracks originating from 
decays and photon conversions. Hadron identification is achieved using the 
time-of-flight detectors, with different technologies in the east 
and west arms, for which the timing resolutions are 130~ps and 95~ps, 
respectively.  Pions and (anti)proton tracks are identified with over 99\% 
purity at momenta up to 3~GeV/$c$~\cite{ppg146,ppg123} in both systems.


Energy deposited at large rapidity in the Au-going direction is measured 
by the towers in the south-side Muon Piston 
Calorimeter~(MPC-S)~\cite{Chiu:2007zy}. The MPC-S comprises 192 towers of 
PbWO$_4$ crystal covering 2$\pi$ in azimuth and $-3.7<\eta<-3.1$ in 
pseudorapidity, with each tower subtending approximately 
$\Delta\eta\times\Delta\phi\approx0.12\times0.18$. Over 95\% of the energy 
detected in the MPC is from photons, which are primarily produced in the 
decays of $\pi^0$ and $\eta$ mesons. Photons are well localized, as 
each will deposit over 90\% of its energy into one tower if it hits the 
tower's center. To avoid the background from noncollision noise sources ($\sim$~75~MeV)
and cut out the deposits by minimum ionization particles ($\sim$~245~MeV), 
we select towers with deposited energy $E_{{\rm Tower}}>3$~GeV.


We first examine the long-range azimuthal angular correlation of pairs 
consisting of one track in the central arm and one tower in the MPC-S. 
Because the towers are mainly fired by photons, and the azimuthal extent 
of each energy deposition is much smaller than the size of azimuthal 
angular correlation from jets or elliptic flow, these track-tower pair 
correlations will be good proxies for hadron-photon correlations without 
attempting to reconstruct individual photon showers. We construct the 
signal distribution $S(\Delta\phi,p_T)$ of track-tower pairs over relative 
azimuthal angle 
$\Delta\phi \equiv \phi_{{\rm Track}} - \phi_{{\rm Tower}}$, each with 
weight $w_{{\rm tower}}$, in bins of track transverse momentum $p_{T}$.

\begin{equation}
  S(\Delta\phi,p_{T})=
  \frac{ d(w_{{\rm Tower}} N^{{\rm Track}(p_{T}){\rm -Tower}}_{{\rm Same \; event}}) }{ d\Delta\phi}
\label{eq31}
\end{equation}
Here $\phi_{{\rm Track}}$ is the azimuth of the track as it 
leaves the primary vertex, $\phi_{{\rm Tower}}$ is the azimuth of the 
center of the calorimeter tower. The $w_{{\rm Tower}}$ is chosen as the 
tower's transverse energy~$E_T=E_{{\rm Tower}} \sin{(\theta_{{\rm 
Tower}})}$. Because the calorimeter is operating in a linear regime the overall 
$E_{T}$ pattern on each event will simply be the sum of the patterns from each 
impinging particle, so we expect no distortion effect due to occupancy. To correct for 
the nonuniform PHENIX azimuthal acceptance in the central arm tracking 
system, we then construct the corresponding ``mixed-event'' distribution 
$M(\Delta\phi, p_{T})$ over track-tower pairs, where the tracks and tower 
signals are from different events in the same centrality and vertex 
position class. We then construct the normalized correlation function

\begin{equation}
  C(\Delta\phi,p_{T}) =
          \frac{S(\Delta\phi,p_{T})}{M(\Delta\phi,p_{T})} \:
          \frac{\int_{0}^{2\pi} M(\Delta\phi,p_{T}) \, d\Delta\phi}{\int_{0}^{2\pi} S(\Delta\phi,p_{T}) \, d\Delta\phi}
  \label{eq:def_corr_function}
\end{equation}
whose shape is proportional to the true pairs distribution over $\Delta\phi$.

\begin{figure}[tb]
  \includegraphics[width=1.0\linewidth]{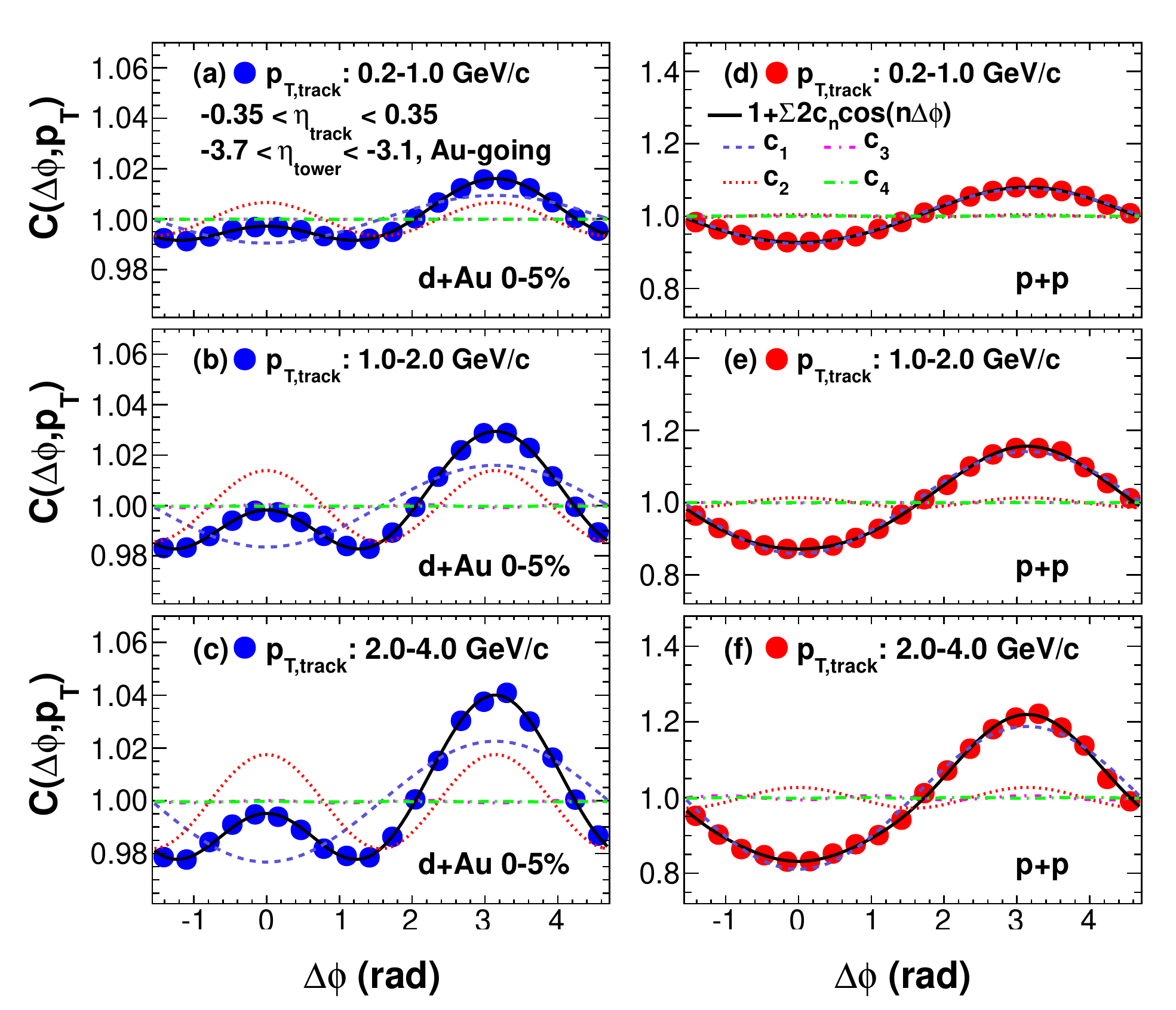}
  \caption{
The azimuthal correlation functions $C(\Delta\phi,p_{T})$, as defined in 
Eq.~\ref{eq:def_corr_function}, for track-tower pairs with different track 
$p_{T}$ selections in 0\%--5\% central $d$$+$Au collisions (left) and 
minimum bias $p$$+$$p$ collisions (right) at \sqsn~=~200~GeV. From top to 
bottom, the track $p_T$ bins are 0.2--1.0~GeV/$c$, 1.0\%--2.0~GeV/$c$ and 
2.0\%--4.0~GeV/$c$. The pairs are formed between charged tracks measured 
in the PHENIX central arms at $|\eta|<0.35$ and towers in the MPC-S 
calorimeter ($-3.7<\eta<-3.1$, Au-going). A near-side peak is observed in 
the central $d$$+$Au which is not seen in minimum bias $p$$+$$p$. Each 
correlation function is fit with a four-term Fourier cosine expansion; the 
individual components $n=1$ to $n=4$ are drawn on each panel, together 
with the fit function sum.
}
  \label{fig:corr_functions}
\end{figure}

Figure~\ref{fig:corr_functions} shows the correlation functions
$C(\Delta\phi,p_{T})$ for different $p_T$ bins, for the 5\%
most central $d$$+$Au collisions and for minimum bias $p$$+$$p$
collisions. Central $d$$+$Au collisions show a visible
enhancement of near-side pairs, producing a local maximum in the
distribution at $\Delta\phi \sim 0$, which is not seen in the
$p$$+$$p$ data. We analyze the distributions by fitting each
$C(\Delta\phi,p_{T})$ to a four-term Fourier cosine expansion,
$f(\Delta\phi) = 1 + \sum_{n=1}^{4} \, 2c_{n}(p_{T}) \, \cos ( n \,
\Delta\phi )$; the sum function and each individual cosine
component are plotted in Fig.~\ref{fig:corr_functions} for each
distribution. We observe that the $p$$+$$p$ distribution shape is
described almost entirely by the dipole term $\cos(\Delta\phi)$, as expected
generically by transverse momentum conservation, via processes such as dijet 
production or soft string fragmentation;
The shape in central $d$$+$Au exhibits both dipole and quadrupole
$\cos(2\Delta\phi)$ terms with similar magnitudes. 
Both $c_3$ and
$c_4$ are found to be $\approx$0, as shown in
Fig.~\ref{fig:corr_functions}.

%
%

\begin{figure}[htb]
  \includegraphics[width=1.0\linewidth]{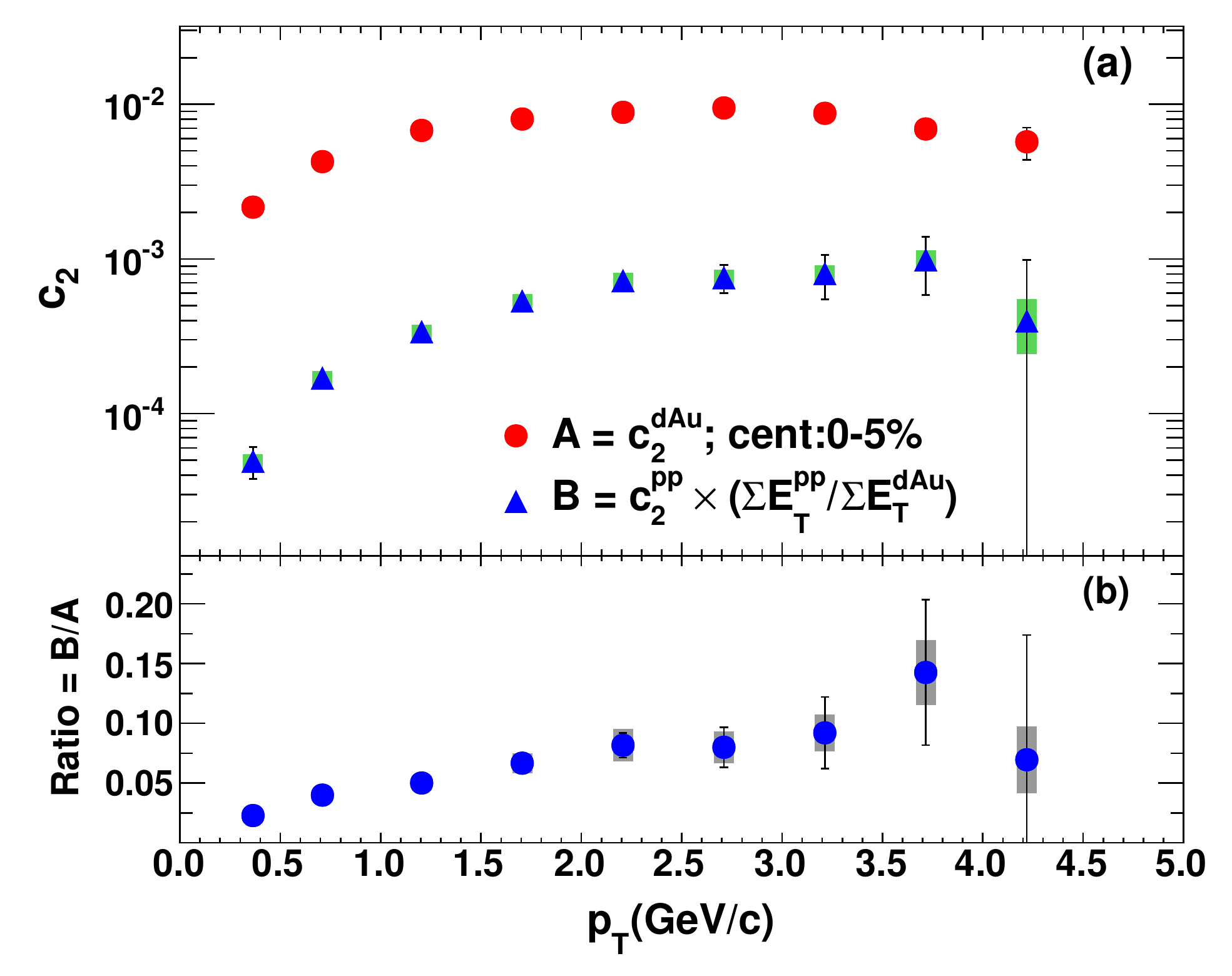}
  \caption{
Panel (a) shows $c_2(p_T)$ for track-tower pairs from 0\%--5\% 
$d$$+$Au collisions and $c_2(p_T)$ for pairs in minimum bias 
$p$$+$$p$ collisions times the dilution factor 
$(\Sigma{E_T}^{pp}/\Sigma{E_T}^{dAu})$.
Panel (b) shows their ratio, indicating that 
the contribution to the $c_2$ amplitude in $d$$+$Au from elementary 
processes present in $p$$+$$p$ are small, only a few percent at low 
$p_T$ and rising to only 10\% by 4.5~GeV/$c$.  Both statistical (bar) and 
systematic (band) uncertainties are shown.
}
\label{fig2}
\end{figure}

Figure~\ref{fig2} shows the fitted $c_{2}$ parameters from the $d$$+$Au 
and $p$$+$$p$ with both statistical and systematic uncertainties. We 
estimate contributions to systematic uncertainties from two main 
sources: (1)~tracking backgrounds from weak decays and photon conversions 
and (2)~multiple collisions in a bunch crossing 
(pile-up) in $d$$+$Au collisions. We estimate the tracking background 
contribution by reducing the spatial matching windows in the third layer 
of the PC from 3$\sigma$ to 2$\sigma$, and find that the change is less than 
2\% fractionally in $c_{2}$. To study the pile-up effect in $d$$+$Au collisions we 
separate the $d$$+$Au data set into two groups, one from a period with 
lower luminosity and the other with the higher luminosity. The 
corresponding pile-up event fractions in central $d$$+$Au are 3.5\% and 7.0\%, 
respectively. The $c_{2}^{dAu}$ in the lower luminosity data set is around 
5\% higher than that in higher luminosity across all $p_T$. The average pile-up 
fraction for the total data sample is around 4\%--5\% and a systematic 
uncertainty around 
10\% is assigned to cover this effect. Additionally, we compare 
$c_{2}^{pp}$ results for $p$$+$$p$ data taken in the 2008 and 2009 running 
periods, and see a difference of less than 5\% for $p_T <$ 1~GeV/$c$, 
increasing to 15\% for $p_T >$ 3~GeV/$c$. 
To characterize biases that might arise because
the tower energy and centrality are measured in the same rapidity range,
we have compared results obtained using two different detectors in the Au-going 
direction to define the event centrality: ($i$) the reaction-plane 
detector ($-2.8<\eta<-1.0$)~\cite{RxNP:2010cc} and ($ii$) the ZDC 
($\eta<-6.5$)~\cite{Baltz19981}.  The $c_2$ values obtained in the 
two cases differ by 6\% from those reported here.


Some portion of the correlation quadrupole strength $c_{2}$ in the 
$d$$+$Au data could be due to elementary processes such as dijet 
fragmentation (mainly from away side) and resonance decays. We can 
estimate the effect of
such processes under the assumptions that (i)~all correlations 
present in minimum bias $p$$+$$p$ collisions are due to elementary 
processes, and (ii)~those same processes occur in the measured $d$$+$Au 
system as a simple superposition of several nucleon-nucleon collisions. 
In this case, we would
expect the contribution from elementary processes to be equal to the 
$c_2^{pp}(p_{T})$ but diluted by the increase in particle 
multiplicity between $p$$+$$p$ and $d$$+$Au, if the number of 
elementary processes is proportional to the multiplicity of the 
other particle used in pair correlations (see also the ``scalar 
product method'', as in~\cite{adams:2004ja, Adams2005:aj}). We 
estimate the ratio of the $p$$+$$p$ to $d$$+$Au general 
multiplicities by measuring the ratio of the total transverse energy 
$\sum E_{T}$ seen in the MPC-S calorimeter in $p$$+$$p$ versus 
$d$$+$Au events, which we find to be approximately $1/(17.9\pm0.35)$ 
and only weakly dependent on the track $p_T$($\le 2\%$).  We can 
then separate $c_2^{dAu}(p_{T})$ into elementary and nonelementary 
components:
\begin{eqnarray}
c_{2}^{dAu}(p_{T}) & = & c_2^{{\rm Non-elem.}}(p_{T}) +
        c_2^{{\rm Elem.}}(p_{T}) \nonumber \\
 & \approx & c_2^{{\rm Non-elem.}}(p_{T}) +
        c_2^{pp}(p_{T}) \frac{ \Sigma{E_{T}^{pp}} }{
   \Sigma{E_{T}^{dAu}}  }
\label{eq:c2_collective_noncollective}
\end{eqnarray}

The ratio in Fig.~\ref{fig2}(b) shows that the contribution to 
$c_{2}^{dAu}$ from elementary processes is indeed small, ranging from a 
few percent at the lowest $p_{T}$ to around 10\% at the highest $p_{T}$, and no more than 
13\% with the other centrality selections mentioned above.
The presence of the near-side peak in the pairs distribution in the central $d$$+$Au system is reproduced in some physics model calculations. The formation of a 
medium that evolves hydrodynamically is one such 
possibility~\cite{Bozek:2013uha,Bzdak:2013zma,Qin:2013bha}, but processes 
such as initial state gluon 
saturation~\cite{Dusling:2012wy,Dusling:2013oia} could also create such an 
effect.


To quantitatively address the physics of this near-side peak and compare with detailed hydro-dynamics calculations,
the $v_2$ of charged hadrons, pions, and (anti)protons at midrapidity is measured via event plane method~\cite{Poskanzer:1998yz}. 
The $v_2$ is measured as $v_{2}(p_{T}) 
= \langle\cos 2 (\phi^{{\rm Particle}} - \Psi_{2}^{{\rm Obs}} ) 
\rangle/{\rm Res}(\Psi_{2}^{{\rm Obs}})$, where 
the average is over particles in the $p_{T}$ bin and over events. The 
second order event plane direction $\Psi_{2}^{{\rm Obs}}$ is determined 
using the MPC-S (Au-going).  The study of correlation strength as above 
indicates that the elementary-process contribution to the event plane 
$v_{2}$ result is similarly small, less than 10\% fractionally out to $p_T 
=$ 4.5~GeV/$c$. The event plane resolution ${\rm Res}(\Psi_{2}^{{\rm 
Obs}})$ ($\sim0.151\pm0.003$) is calculated through the standard three subevents 
method~\cite{Poskanzer:1998yz,Afanasiev:2009wq}, with the other two event 
planes being (i)~the second order event plane determined from central-arm 
tracks, restricted to low $p_T$ (0.2~GeV/$c$ $<p_T<$ 2.0~GeV/$c$) to 
minimize contribution from jet fragments; and (ii)~the first order event 
plane measured with spectator neutrons in the shower-maximum detector on 
the Au-going side ($\eta<$ -6.5)~\cite{Baltz19981,Afanasiev:2009wq}. 
The systematic uncertainties on the $v_2$ of charged 
hadrons are mainly from the tracking background(2\%) and pile-up effects(5\%), as 
described above, and also from the difference in $v_2$ from different 
event plane determinations. To estimate the systematic uncertainty of the 
latter we compare the $v_2$ extracted with the MPC-S event plane with that 
using the south (Au-going) beam-beam counter, and the two measurements of 
$v_{2}$ are consistent to within 5\%. The difference for $v_2$ from the different 
centrality determinations as discussed previously is less than 3\%.

%
%
%

\begin{figure}[tb]
  \includegraphics[width=1.0\linewidth]{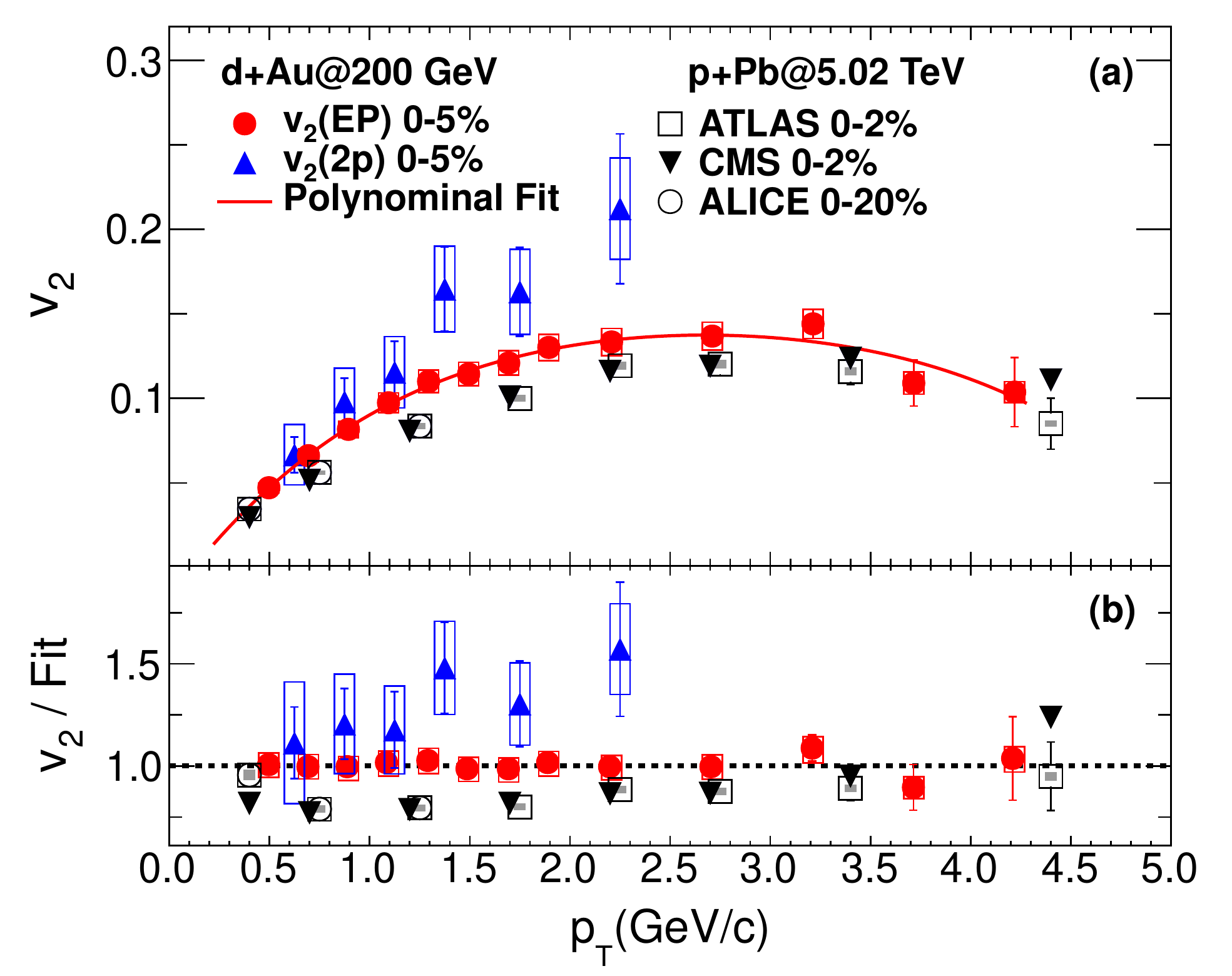}
  \caption{
Measured $v_2(EP)$ for midrapidity charged tracks in 0\%--5\% central 
$d$$+$Au at \sqsn~=~200~GeV using the event plane method in Panel (a).  
Also shown are $v_2$ measured in central $p$$+$Pb collisions at 
\sqsn~=~5.02~TeV~\cite{Aad:2012,Abelev:2012cya,CMS:2013cm}, and our prior 
measurements with two particle correlations ($v_2(2p)$) for $d$$+$Au 
collisions ~\cite{Adare:2013piz}.  A polynomial fit to 
the current measurement and the ratios of experimental 
values to the fit are shown in the panel (b).
}
\label{fig3}
\end{figure}


The $v_{2}$ of charged hadrons for 0\%--5\% central $d$$+$Au events with 
event plane methods are shown in Fig.~\ref{fig3}(a) as $v_{2}(EP)$ for 
$p_{T}$ up to 4.5~GeV/$c$, along with a polynomial fit through the points.  
Also shown are our earlier measurement with two particle correlations 
($v_2(2p)$) and the $v_2$ measured in the central $p$$+$Pb collisions at 
LHC. Figure~\ref{fig3}(b) shows the ratios of all of these measurements 
divided by the fitting results. The $v_2$ from our prior measurements, 
with subtraction of peripheral data to reduce jet contributions, 
exceed the current measurement; differences range from about 15\% at 
$p_{T} =$ 1.0~GeV/$c$ and increases to about 50\% at $p_{T} =$ 
2.2~GeV/$c$. The difference is about 1.5 $\sigma$ for the top three points 
with the largest deviations from the fit. It may be due to different 
jet-like correlation being present in central and peripheral 
collisions~\cite{STAR:2014du}. The present measurement, 
without peripheral subtraction, is performed with 
$|\Delta\eta| >$ 2.75, far away from the near-side main jet peak. The 
contribution from jet, which includes both near and away-side, has been 
found to be less than 10\% from the study of $c_2$ shown in 
Fig.~\ref{fig2}.  Even if there is a 30\% enhancement of jet-like 
correlation from $p$+$p$ to central $d$$+$Au collisions, it will only 
raise from 10\% to 13\% our estimate of the jet-like contribution to the 
$v_2$ in central $d$$+$Au collisions. 
The present $v_2$ measurement is closer to that of $p$$+$Pb 
collisions~\cite{Aad:2012,Abelev:2012cya,CMS:2013cm}. It is about 20\% 
higher than that of $p$$+$Pb at $p_T=1$~GeV/$c$, and the difference 
decreases to a few percent at $p_{T} >$ 2.0~GeV/$c$.

%
%
%

\begin{figure}[htb]
  \includegraphics[width=1.0\linewidth]{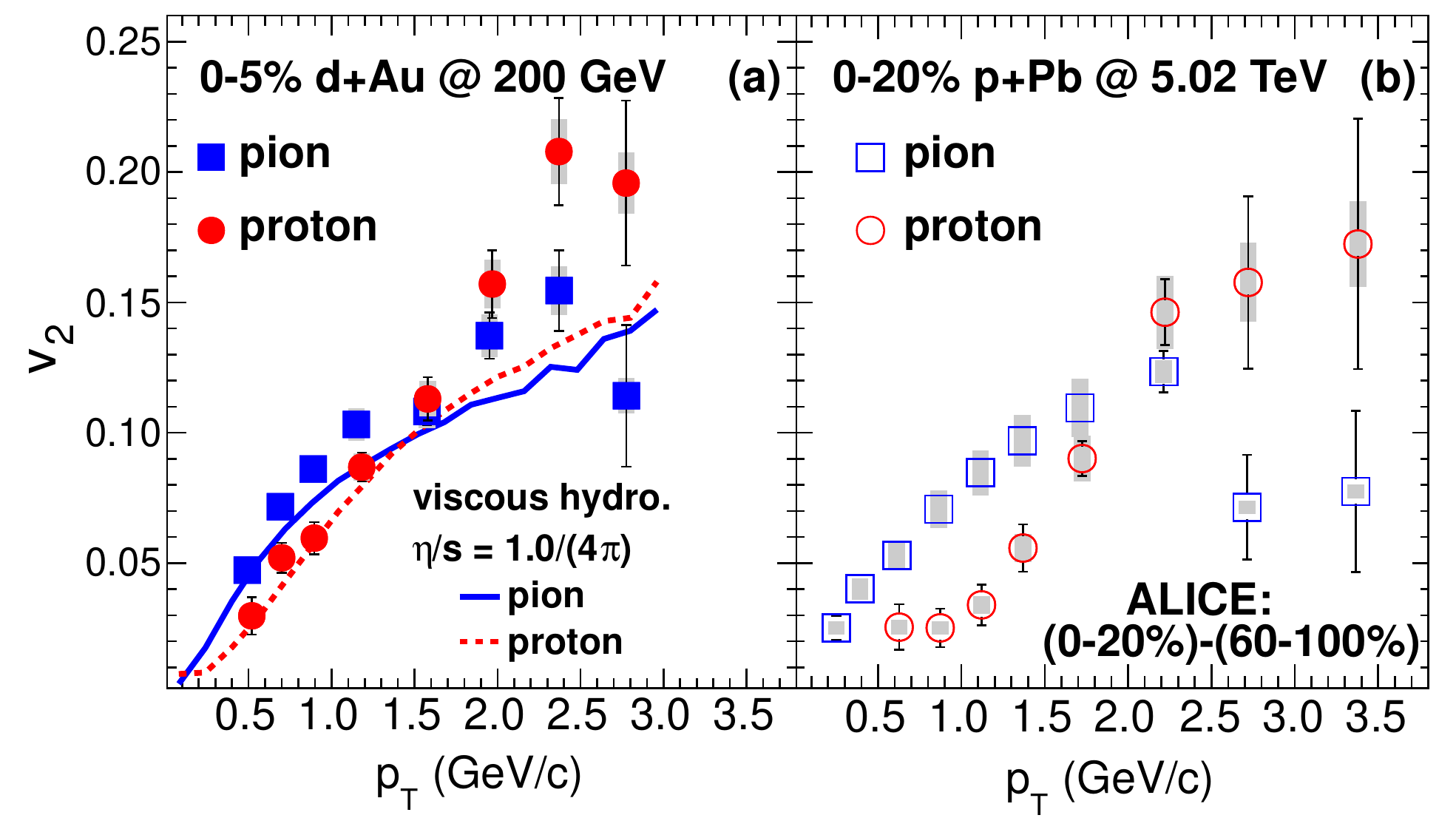}
  \caption{
Measured $v_2(p_T)$ for identified pions and (anti)protons, each charged 
combined, in 0\%--5\% central $d$$+$Au collisions at RHIC. In panel~(a) 
the data are compared with the calculation from a viscous hydrodynamic 
model~\cite{Nagle:2013ei,Paul:privatecomm,PhysRevC.78.034915}, and in 
panel~(b) the $v_2$ data for pions and protons in 0\%--20\% central 
$p$$+$Pb collisions at LHC are shown for comparison~\cite{ABELEV:2013wsa}, they are 
measured from pair correlations with a peripheral event yield subtraction
}
\label{fig4}
\end{figure}

Figure~\ref{fig4} shows the midrapidity $v_{2}(p_{T})$ for identified 
charged pions and (anti)protons, with charge signs combined for each 
species, up to $p_{T}=3$~GeV/$c$ using the event plane method; the 
systematic uncertainties are the same as for inclusive charged hadrons. 
A distinctive mass-splitting can be seen. The pion $v_2$ is higher than 
the proton's for $p_{T}<1.5$~GeV/$c$, as has been seen universally in 
heavy-ion collisions at 
RHIC~\cite{Adler:2003kt,Adams:2003am,Adare:2006ti,Afanasiev:2007tv,Abelev:2007qg,Abelev:2008ed}. 
Figure~\ref{fig4}(a) also 
shows calculations of viscous hydrodynamics with Glauber initial conditions 
starting at $\tau=0.5$~fm/$c$ with $\eta/s=1.0/(4\pi)$,
followed by a hadronic 
cascade~\cite{Nagle:2013ei,Paul:privatecomm,PhysRevC.78.034915}.
The splitting at lower $p_{T}$ is also seen in the calculation.
The identified particle $v_2$ in 0\%--20\% $p$$+$Pb 
collisions are shown in Fig.~\ref{fig4}(b) for 
comparison~\cite{ABELEV:2013wsa}. The magnitude of the mass-splitting in 
RHIC $d$$+$Au is smaller than that seen in LHC $p$$+$Pb, which could be 
an indicator of stronger radial flow in the higher energy collisions~\cite{Chun:2011cs}.


We have presented measurements of long-range azimuthal correlations 
between particles at midrapidity and at backward rapidity (Au-going 
direction) in 0\%--5\% central $d$$+$Au collisions at \sqsn=~200~GeV.  We 
find a localized near-side azimuthal angular correlation in these collisions for 
pairs across $|\Delta\eta| >$~2.75 which is not apparent in minimum bias 
$p$$+$$p$ collisions at the same collision energy. The anisotropy strength 
$v_2$ is measured for midrapidity particles with respect to a event 
plane determined from a region separated by the same pseudorapidity 
interval. The $v_2$ values are qualitatively similar to those observed in central 
$p$$+$Pb collisions at \sqsn=~5.02~TeV. 
The $v_2$ for identified pions and (anti)protons at midrapidity exhibit a mass ordering,
qualitatively similar to observations in relativistic heavy-ion collisions.
This ordering can be described by a viscous hydrodynamic 
model, where they are believed to reflect radial flow in hydrodynamics. 
The magnitude of mass-splitting in $v_2(p_t)$ is found to be smaller in 
$d$$+$Au collisions in comparison to $p$$+$Pb collisions at higher 
energies, possibly indicating smaller radial flow in $d$$+$Au at 
\sqsn=~200~GeV.




We thank the staff of the Collider-Accelerator and Physics
Departments at Brookhaven National Laboratory and the staff of
the other PHENIX participating institutions for their vital
contributions.  We acknowledge support from the 
Office of Nuclear Physics in the
Office of Science of the Department of Energy, 
the National Science Foundation, 
Abilene Christian University Research Council, 
Research Foundation of SUNY, and 
Dean of the College of Arts and Sciences, Vanderbilt University (U.S.A),
Ministry of Education, Culture, Sports, Science, and Technology
and the Japan Society for the Promotion of Science (Japan),
Conselho Nacional de Desenvolvimento Cient\'{\i}fico e
Tecnol{\'o}gico and Funda\c c{\~a}o de Amparo {\`a} Pesquisa do
Estado de S{\~a}o Paulo (Brazil),
Natural Science Foundation of China (P.~R.~China),
Ministry of Education, Youth and Sports (Czech Republic),
Centre National de la Recherche Scientifique, Commissariat
{\`a} l'{\'E}nergie Atomique, and Institut National de Physique
Nucl{\'e}aire et de Physique des Particules (France),
Bundesministerium f\"ur Bildung und Forschung, Deutscher
Akademischer Austausch Dienst, and Alexander von Humboldt Stiftung (Germany),
Hungarian National Science Fund, OTKA (Hungary), 
Department of Atomic Energy and Department of Science and Technology (India), 
Israel Science Foundation (Israel), 
National Research Foundation of Korea of the Ministry of Science,
ICT, and Future Planning (Korea),
Physics Department, Lahore University of Management Sciences (Pakistan),
Ministry of Education and Science, Russian Academy of Sciences,
Federal Agency of Atomic Energy (Russia),
VR and Wallenberg Foundation (Sweden), 
the U.S. Civilian Research and Development Foundation for the
Independent States of the Former Soviet Union, 
the Hungarian American Enterprise Scholarship Fund,
the US-Hungarian Fulbright Foundation for Educational Exchange,
and the US-Israel Binational Science Foundation.



%
 
\end{document}